\documentclass[12pt]{article}
\usepackage{cite}
\usepackage{amsmath, amsthm, amssymb,slashed,graphicx,epsfig}

\newcommand{\smallfrac}[2] {\mbox{$\frac{#1}{#2}$}}
\newcommand {\slsh} [1] {\not{\hbox{\kern-2pt${#1}$}}}

\newcommand{\gsim}{\lower.7ex\hbox{$\;\stackrel{\textstyle>}{\sim}\;$}}
\newcommand{\lsim}{\lower.7ex\hbox{$\;\stackrel{\textstyle<}{\sim}\;$}}
\newcommand {\beq} {\begin{equation}}
\newcommand {\eeq} {\end{equation}}
\newcommand {\beqn}{\begin{eqnarray}}
\newcommand {\eeqn} {\end{eqnarray}}
\newcommand{\bea}{\begin{eqnarray}}
\newcommand{\eea}{\end{eqnarray}}

\def\N{{\mathcal N}}

\def\be{\begin{eqnarray}}
\def\ee{\end{eqnarray}}

\def\lae{\mathrel{\mathop{\smash{\lower .5 ex \hbox{$\stackrel<\sim$}}}}}
\def\lae{\mathrel{\mathop{\smash{\lower .5 ex \hbox{$\stackrel>\sim$}}}}}

\def\beqn{\begin{eqnarray}}
\def\eeqn{\end{eqnarray}}

\def\ba{\beq\new\begin{array}{c}}
\def\ea{\end{array}\eeq}
\def\be{\ba}
\def\ee{\ea}

\begin{document}

\begin{titlepage}

\begin{flushright}
FTPI-MINN-09/25, UMN-TH-2805/09\\
July 20, 2009
\end{flushright}

\vskip 0.4in
\begin{center}
{\bf\Large{Quantum Fusion of Domain Walls 
\\[2mm]
with Fluxes}}\vskip0cm
\vskip 0.5cm  {\bf\large{S.~Bolognesi$^a$,  M.~Shifman$^{a,b}$,  and  M.B.~Voloshin$^{a,c}$}} \vskip
0.05in $^a${\small{ \textit{William I. Fine Theoretical Physics Institute, University of Minnesota, } \\ {\textit{116 Church St. S.E., Minneapolis, MN 55455, USA}}}}
\vskip
0.05in 
$^b${\it Institut de Physique Th\'eorique, CEA Saclay, 91191 Gif-sur-Yvette C\'edex, France}\\
\vskip
0.05in
$^c${\small{ \textit{Institute of Theoretical and Experimental Physics,117218, Moscow, Russia } }}
\end{center}
\vskip 0.5in

\baselineskip 10pt
%

\begin{abstract}

We study how 
fluxes on the domain wall world volume modify
 quantum fusion of two distant parallel domain walls into a composite wall.
The elementary wall fluxes can be separated into parallel and antiparallel components. 
The parallel component affects neither the binding energy nor   the process of quantum merger. 
The antiparallel fluxes, instead, increase the binding energy and, against   naive expectations, 
suppress
quantum fusion. 
In the small flux limit we explicitly find  the bounce solution and  the
fusion rate as a function of the flux. We argue that at large (antiparallel) fluxes there exists
 a critical value of the flux (versus the difference in the wall tensions), which switches off quantum fusion altogether.
This phenomenon of flux-related wall stabilization is rather peculiar: it is  unrelated to any conserved quantity. Our consideration of the  flux-related wall stabilization
is based on substantiated arguments that fall short of complete proof.

\end{abstract}

\end{titlepage}
\vfill\eject

\section{Introduction}
\label{intro}

In our previous paper \cite{Bolognesi:2009kp} we showed how to describe quantum fusion of 
two parallel elementary domain walls with tension $T_1$ into a composite wall with tension $T_2$, 
with a binding energy, i.e.  $T_2 < 2 T_1$.
The  distance $d$ between the elementary walls is assumed to be
much larger than the wall thickness.
An illustrative example in which composite domain walls have a binding energy, 
and our calculation can be applied, is that of the $k$-walls in $\N=1$ super-Yang--Mills  theory \cite{Dvali:1999ht}.

Many microscopic theories supporting domain walls allow one to introduce constant 
magnetic fluxes inside the walls 
\cite{Shifman:2002jm,Auzzi:2008zd}. 
In these cases the corresponding wall world-volume theory is $(2+1)$-dimensional QED,
\beq
{\cal L}_{2+1} = -\frac{1}{4 {e_{2+1}}^2} {F_{\mu \nu}}^2 \ .
\eeq
(In what follows we will omit the subscript 2+1, as the only electric charge
that will appear below is that of the $(2+1)$-dimensional theory.)
It is well known \cite{Polyakov:1976fu} that in $2+1$ dimensions the electromagnetic 
field can be dualized into a compact scalar field $\sigma$, defined mod $2\pi$,
\beq
F_{\mu \nu} = \frac{e^2}{2\pi} \varepsilon_{\mu\nu\rho} \, \partial^{\rho} \sigma \,.
\eeq
The Lagrangian takes the form
\beq
{\cal L}_{2+1} =\frac{e^2}{8\pi^2} \left(\partial_{\mu} \sigma\right)^2 \, .
\eeq
The field $\sigma$  linearly depending on a spatial coordinate $x^i$
\beq
\label{flux}
\sigma = {n^\mu} {x^\mu} \, , \qquad n_y \equiv n\,,
\eeq
with all other components of $n^\mu$ vanishing,
describes a constant electric field on the wall world volume, which is in one-to-one 
correspondence with the magnetic flux trapped inside the wall in the 
bulk description \cite{Auzzi:2008zd}.

In this paper we consider the impact of  possible  fluxes  on the wall quantum fusion. 
For parallel fluxes, the binding energy and the fusion rate remain unchanged. Therefore,
the question we focus on is the impact of
antiparallel fluxes. In this case the flux contribution effectively 
increases the tension of
the elementary walls. Since on the composite wall the flux vanishes,
$T_2$ stays intact. As a result, effectively,
\beq
\Delta T\equiv 2T_1-T_2
\eeq
increases which, 
at first sight,  should entail an enhancement of the fusion rate.

We will show that, in fact, it is the opposite tendency that prevails:
switching on antiparallel fluxes on the elementary walls increases the bounce action and, hence,
{\em suppresses} the fusion rate.

Our consideration proceeds in two stages. First, we consider the problem
 in the limit $e^2n^2 \ll \Delta T$. In this limit the bounce solution can be explicitly determined,
 and its action analytically calculated. The fusion rate obtained in the no-flux problem \cite{Bolognesi:2009kp}
 \beq
 \Gamma \propto e^{-S_B}\,,\qquad S_{B} =  \frac{\pi}{3}\,\, { T_1 d^3}\, \sqrt{\frac{T_1}{\Delta T}} 
 \eeq
 gets modified in a rather minimal way,
 \beq
 \left(S_B\right)_{n\neq 0} =\frac{\pi}{3}\, T_1\, d^3\,\sqrt{\frac{T_1}{\Delta T - \frac{e^2}{4\pi^2}\,n^2}}\,\,.
 \label{facev}
 \eeq
 The suppression of the fusion rate is obvious in Eq.~(\ref{facev}).
Taking Eq.~(\ref{facev}) at its face value, we would  conclude that
for the fusion to occur
$\Delta T$ must exceed a  critical value,
 $\Delta T_* = \smallfrac{e^2n^2}{4\pi^2}$. At this critical point 
 the fusion rate vanishes; and at $\Delta T < \Delta T_*$ 
 the elementary wall fusion through quantum tunneling  becomes impossible
since the Euclidean bounce configuration no longer exists.
 
 Equation~(\ref{facev}), literally speaking,
becomes invalid at $\Delta T_* \sim \smallfrac{e^2n^2}{4\pi^2}$. 
Its derivation, to be presented below, is based on the small flux assumption.
This assumption is relaxed at the second stage.
We argue that  the conclusion survives at a qualitative level:
 $ \left(S_B\right)_{n\neq 0}\to \infty $ at some finite positive value of $\Delta T = \Delta T_*$,
 \beq
 \Delta T_* = {\rm const}\, \cdot \, \frac{e^2n^2}{4\pi^2}\,,
 \eeq
 where the constant appearing on the right-hand side is of order 1.
  We call this phenomenon
 flux-induced stabilization of the wall fusion. The flux-induced stabilization
 is an interesting and rather peculiar phenomenon. 
Usually, when speaking of flux stabilization, we have in mind something related to 
conserved quantities. For example,
 some radius (which can be the size of a soliton or of a cycle on a manifold) 
 can be stabilized by a  flux
 captured inside. Combined application of 
 energy and flux conservation prevents the radius from  shrinking to zero,
 stabilizing  the object under consideration.
In our problem, instead, conservation laws would be perfectly consistent with, and, 
moreover, in favor of
the fusion of the two separated elementary walls. The fusion is prohibited by the absence 
of any finite-action configuration
(bounce)  that could mediate the process. In this sense the situation is similar to that
discovered by Coleman and De Luccia \cite{codelu} who found that gravity-related effects 
suppress the process of false vacuum decay through bubble creation.
It is worth noting that a
suppression of the existence of a bounce was also observed in \cite{GSCH}
from the string/$D$-brane theory side.
The problem  considered in \cite{GSCH} was different (creation of pairs of particles,
charged electrically or magnetically, in external fields), but the bounce
suppression  and, in particular, the existence of a critical electric field, seemingly have a common origin
which can be traced back to \cite{Bachas:1992bh} where the rate of pair production of open bosonic and supersymmetric strings in a constant electric field was discussed. If we continued this parallel, at an intuitive level,
we might conjecture that,
in our problem, when   $n$ reaches its critical value, the bounce solution under consideration breaks into
separate 
pieces.

It should be also mentioned that the process we consider is the fusion of walls due to quantum tunneling. Generally, depending on particular situation, the fusion can proceed classically, e.g. if the walls actually intersect or if they move toward each other. For strictly parallel walls that are exactly at rest a classical fusion is possible due to an exponentially weak attraction between them at large distances. Clearly, the classical fusion crucially depends on the initial conditions. For instance any exponentially weak attraction becomes irrelevant if the walls are slowly moving away from each other, so that the classical fusion never occurs. The behavior of the quantum fusion, considered in the present paper, is different in that the exponential factor in the probability has a smooth dependence on the initial conditions, such as (a small) relative velocity of the walls.

The microscopic theory supporting
domain walls with fluxes, which can be taken as an example, is  a 
straightforward generalization of that 
of Ref.~\cite{Shifman:2002jm}, namely,  $\,\N=2$ SQED, with three matter 
hypermultiplets with masses $m_{1,2,3}$, and a Fayet--Iliopoulos term. There are 
three vacua, according to which hypermultiplet's scalar field $a$ is locked. 
Correspondingly,
there are three domain walls.
We will refer to two walls with smaller tension, say $\langle12\rangle$ and $\langle23\rangle$ 
as to elementary walls. 
Then the wall $\langle13\rangle$ is composite.
The masses $m_{1,2,3}$ are complex parameters. If they are not aligned in the complex plane, 
there is a natural binding energy between the elementary walls. If they are 
aligned, $T_2=2T_1$ and $\Delta T = 0$.

We will assume that $\Delta T  \ll T_1$ so that, instead of the 
full Dirac--Born--Infeld action on the wall world volume, we can limit ourselves to quadratic terms.
Geometry of the problem, our notation and all constraints and limitations
are the same as in
Ref.~\cite{Bolognesi:2009kp}.

\section{Switching on fluxes in the limit \boldmath{$e^2n^2 \ll \Delta T$}}
\label{small}

The existence of the flux (\ref{flux}) entails two consequences. First, the effective wall tension changes.
If, without the fluxes, it was $T_1$, then, with the fluxes switched on 
it becomes
\beq
T_1 \to T_1 + \delta \equiv T_1 +\frac{e^2}{8\pi^2}\, {n}^2 \,.
\label{five}
\eeq
Without loss of generality we can assume that the fluxes
on two elementary walls are the same in the absolute value and {\em anti}parallel. 
Then the flux on the composite wall must vanish. Correspondingly,
the tension of the composite wall $T_2$ remains intact. Note that $n$ has dimension of mass,
and so does the three-dimensional coupling $e^2$.

The second novel element is the necessity of matching of  the $\sigma$ fields 
on the elementary walls at the boundary of the fused domain.

\begin{figure}[h!t]
\epsfxsize=9cm
\centerline{\epsfbox{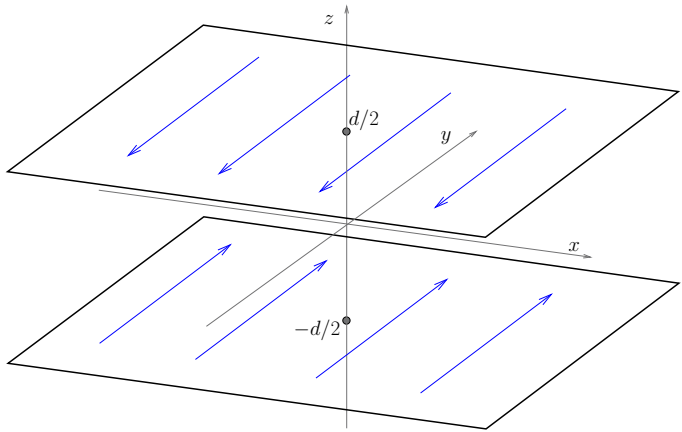}}
\caption{{\footnotesize Geometry of the problem. Two elementary walls at distance $d$ with anti-parallel fluxes.     }}
\label{geometry}
\end{figure}

The geometry of the problem is exhibited in Figure \ref{geometry}: we have two elementary parallel 
walls $\langle12\rangle$ and $\langle23\rangle$ lying in the $x,y$ plane,  at separation $d$ 
in the $z$ direction. There are antiparallel  fluxes on both of 
them.\footnote{If the fluxes are
not antiparallel, say,
 $n_{x, {\langle12\rangle}} =  n_{x,  {\langle23\rangle}}$ and $n_{y, {\langle12\rangle}} = - n_{y, {\langle23\rangle}}$,
the tensions of the two elementary walls are
$
T_{{\langle12\rangle}} = T_{{\langle23\rangle}} = T_1 + \frac{e^2}{8\pi^2} (n_x^2 + n_y^2 )
$
while
the tension of the composite wall  is 
$
T_{{\langle13\rangle}} =  T_2 + \frac{e^2}{2\cdot 8\pi^2} (2 n_x)^2
$. Since the $x$ component matches automatically (and $n_x$ = const on the
bounce), we can include the $n_x^2$
terms in the corresponding tensions. Then the problem reduces to that
with antiparallel fluxes.
Note  two factors. 
The electric coupling $e^2$ of the composite wall is half of that of the elementary walls.
If the fluxes are parallel (i.e. $n_y =0$) the energy stored in the flux has no binding effect. 
Since the $n_x$ component has no effect, apart from an overall shift of the tensions, 
we  set it to zero.}

In this section we will consider misaligned mass parameters $m_i$ in the microscopic
theory, i.e. $2 T_1 > T_2$.
Our task is to find the bounce solution satisfying the following conditions:
(i) In the linear approximation both $z$ and $\sigma$ satisfy the Laplace equation, and, 
in particular, they do not interact,
\beq
\label{laplace}
\Delta z =0 \ , \qquad \Delta \sigma =0 \, .
\eeq
(ii) The boundary conditions are set at infinity in  $\tau,\, x,\, y$ where $\tau $ is  
Euclidean time, and at the boundary of the fused domain at $z=0$. Both $z$ and $\sigma$ must match
at $z=0$, while at infinity
$z\to \pm d/2$ and $\sigma \to \pm n_\mu x^\mu$.  Here 
\beq
x^\mu \equiv\{\tau ,\,x,\, y\}\,.
\eeq
The plus-minus signs refer to the upper and lower walls, respectively.

If 
\beq
e^2n^2\ll \Delta T\ll T_1
\label{conds}
\eeq
the spherical symmetry of the bounce field configuration (i.e. the fact that 
$z(x^\mu) = f \left(\sqrt{x^\mu x^\mu}\right)$)
remains approximately valid, since the $\sigma$ related contribution can be considered as
a small correction and its back reaction ignored.
One can use the multipole expansion for the bounce solution
keeping only the lowest harmonics, i.e. $l=0$ for $z$ and $l=1$ for $\sigma$.
Then,
to  the leading order in this expansion, for the upper brane,
\beq
z= \frac{d}{2}\left(1 - \frac{r_*}{r}\right)\,,\qquad 
\sigma = n_\mu x^\mu \left(1 - \frac{r_*^3}{r^3}\right),
\label{eight}
\eeq
where 
\beq
r \equiv \sqrt{x^\mu x^{\mu }}
\eeq
and $r_*$  is the radius
of the fused domain with the composite wall at $z=0$. Please, note that
the solution (\ref{eight}) satisfies both boundary conditions. At $r=r_*$, on the 
boundary of the composite wall,
$z=0$ and $\sigma = 0$. The value of $r_*$ is to be determined through
extremization (maximization) of the bounce action.
 
Now we have to calculate the bounce action. It consists of a few distinct contributions:
(a) at $r>r_*$ we loose, compared to the two flat walls, due to the fact
$z$ and $\sigma$ nontrivially depend on $x^\mu$; on the other hand,
at $r<r_*$ we gain due to the fact $T_2 < 2T_1 +\frac{e^2}{4\pi^2}n^2$.
The extremal balance between gain and loss is achieved at a critical (extremal) value of $r_*$.

The loss due to the wall curvature (i.e. $z\neq$ const.) is \cite{Bolognesi:2009kp}
\beq
\left(\Delta S_z\right)_{r>r_*} = \pi T_1 r_* d^2\,.
\label{nine}
\eeq
 The flux-related loss (due to $\sigma \neq
\sigma_{\rm asymptotic}$) is
\beq
\left(\Delta S_\sigma \right)_{r>r_*}
= 2\times
 \frac{e^2}{8\pi^2}  \,  \int_{r_*}^{\infty}\,d\Omega \, r^2\,dr\, 
\left[
( \partial \sigma)^2 - {n}^2 \right]
=
\frac{2e^2}{3\pi}\,r_*^3\, n^2\,,
\label{ten}
\eeq
where $d\,\Omega$ presents the angular integration, $\int d\Omega =4\pi$, and the overall factor of 2 is
due to the fact that
both elementary walls are included in  (\ref{ten}). (The same is valid for (\ref{nine}).)

The gain from the domain $r<r_*$ is
\beq
\Delta S_{r<r_*}
\equiv \left(\Delta S_z\right)_{r<r_*} + \left(\Delta S_\sigma \right)_{r<r_*}
 = -\frac{4\pi\,r_*^3}{3}\left(2T_1 +\frac{e^2}{4\pi^2}\,n^2 -T_2\right).
\label{eleven}
\eeq
The total contributions due to $\sigma$ and $z$ are as follows:
\beqn
\Delta S_\sigma 
&=&
 \left(\Delta S_\sigma \right)_{r>r_*} +\left(\Delta S_\sigma \right)_{r<r_*} = \frac{e^2}{3\pi}\,r_*^3\, n^2\,,
\nonumber\\[3mm]
 \Delta S_z &=&
  \left(\Delta S_z\right)_{r>r_*} +\left(\Delta S_z \right)_{r<r_*} =
  \pi T_1 r_* d^2 - \frac{4\pi\,r_*^3}{3}\,\Delta T\,.
  \label{dssdsz}
\eeqn
Note that the $\sigma$-related gain at $r<r_*$
is over-compensated by the loss at $r>r_*$. Equation (\ref{dssdsz})
implies
\beq
\Delta S = \Delta S_z+\Delta S_\sigma  =  \pi T_1 r_* d^2 - \frac{4\pi\,r_*^3}{3}\left(\Delta T
-\frac{e^2}{4\pi^2}\,n^2\right)\,,
\label{dees}
\eeq

Next, to find the critical radius,
we must extremize $\Delta S$
with respect to $r_*$. 
The functional form of $\Delta S $ in (\ref{dees})
is the same as in the problem \cite{Bolognesi:2009kp} with vanishing fluxes.
As a result, the extremal value of the radius of the fused domain 
takes the form
\beq
r_* = \frac{d}{2} \,\sqrt{\frac{T_1}{\Delta T - \frac{e^2\,n^2}{4\pi^2}}} \,,
\eeq 
while  the bounce action is
\beq
S_B =\frac{\pi}{3}\, T_1\, d^3\,\sqrt{\frac{T_1}{\Delta T - \frac{e^2\,n^2}{4\pi^2}}}
 \,.
\label{fourteen}
\eeq
As usual, the fusion rate per unit time per unit area of the wall
is proportional to
\beq
\Gamma \sim e^{-S_B}\,.
\eeq

A few explanatory comments are in order here.
Equation (\ref{fourteen}) presents the {\em maximal} value of $\Delta S$ as a function of $r_*$
(as opposed to minimal),
in full accordance with the fact the fusion process under consideration
is that of quantum instability of two flat elementary walls. 
The flux-related contribution suppresses the decay rate.
If we formally extrapolate the result to $\Delta T = \frac{e^2\,n^2}{4\pi^2}$,
at this point $S_B\to \infty$ and the suppression becomes absolute.
There is no wall fusion below this point. However,
Eq. (\ref{fourteen}) is valid only at small $e^2n^2/\Delta T$, to the leading order in 
this parameter. This is due to the fact that we used 
the spherically symmetric ansatz (\ref{eight}) for $z(\tau, x, y)$.  
When $e^2n^2/\Delta T\sim 1$
this ansatz is no longer justified.
Deviations from sphericity
will be discussed in the next section.

\section{The general case: \boldmath{$e^2n^2 \lsim \Delta T$}}
\label{generic}

In supersymmetric theories with critical (BPS saturated) walls
the degenerate situation 
\beq
T_2=2T_1
\label{degen}
\eeq
is not uncommon. Then, if
there are no fluxes the two elementary 
domain walls at rest at separation $d$ present an absolutely stable configuration.
Quantum fusion is impossible.
If we switch on antiparallel fluxes on these walls, according to 
(\ref{five}), the tension of the elementary walls increases while that
of the composite wall stays intact, i.e. effectively
$2T_1$ becomes larger than $T_2$. Therefore, one might suspect that
switching on fluxes induces quantum fusion  in the degenerate case. 
In fact, as was shown above, the tendency  is just
opposite.  Equation (\ref{fourteen}) suggests that there may exist
a critical  value of $\Delta T /e^2 n^2$ below which
the wall fusion through tunneling becomes impossible. 

To verify this hypothesis we
have to move away from the small flux limit, i.e. relax the first condition in (\ref{conds}).
The spherical approximation  valid for small fluxes is not
expected to be applicable at $e^2n^2/4\pi^2 \sim \Delta T$. Indeed, at
$e^2n^2/4\pi^2 \sim \Delta T$ the dipole contribution due to $\sigma$ becomes
important  in the determination of the bounce shape. It feeds back into
the solution for  $z(\tau, x,y)$,
generating  angular momentum $l=2$ in the $z$ profile. This, in turn,
 triggers $l=3$ harmonics in the $\sigma$ solution, and so on. 
 All terms in the multipole expansion enter the game.
 We will not be able to find the exact answer for the bounce configuration
 in this case. However, certain predictions are still possible. 
 
Let us begin by discussing general lessons we can abstract from Sect.~\ref{small}.
The bounce solution has the following features. There is a certain domain ${\mathcal M}$
(which includes the origin)  inside which the two elementary walls are merged;
on the boundary of this domain $\partial {\mathcal M}$ and inside it
$z=0$ and  $\sigma =0$. Outside ${\mathcal M}$  both $z$ and $\sigma$ satisfy the Laplace equations (\ref{laplace}) with the  boundary conditions at $\infty$ 
\beq
z\to \pm d/2\,,\qquad\sigma \to \pm n^\mu x^\mu \,,\qquad r\to\infty\,,
\label{24}
\eeq
where $n^\mu = \{0,0,n\}$. A typical linear dimension of ${\mathcal M}$
$$ \ell_{\mathcal M}\gg d\,.$$
In addition to these general lessons  from Sect.~\ref{small} we should add 
a particular lesson: the absolute value of the flux-related loss
$\left|\left(\Delta S_\sigma\right)_{r>r_*}\right|$ is larger than
that of the flux related gain $\left|\left(\Delta S_\sigma\right)_{r<r_*}\right|$.
Below we will argue that this crucial feature 
bears a more general nature than sphericity.

As was mentioned, beyond the small flux limit,
spherical symmetry is lost (although the axial symmetry survives);
in particular, $\partial {\mathcal M}$ is no longer $S_2$. It is worth emphasizing that
the equations are still linear (see Eq.~(\ref{laplace})); 
the coupling between $z$ and $\sigma$ is realized through the shape 
of $\partial {\mathcal M}$.
The condition of the tension balance at the boundary $\partial {\mathcal M}$ is
\beq
\label{balance}
\left( 2T_1 + \frac{{e}^2}{4\pi^2} (\partial \sigma )^2 \right) \frac{1}{ \sqrt{1+(\partial z)^2} }= T_2 \,  ,
\eeq
where on the left-hand side
 we have the tension of the two external branes multiplied by the cosine of the angle at which they merge, while on the right-hand side the tension of the composite brane.
This information seems to be sufficient to determine, at least qualitatively, the shape of the domain $ {\mathcal M}$.
\begin{figure}[h!t]
\epsfxsize=3.5cm
\centerline{\epsfbox{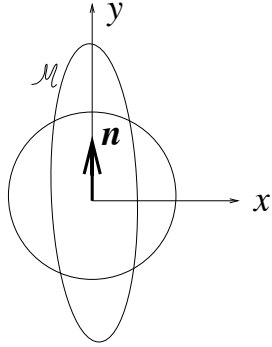}}
\caption{{\footnotesize A slice of the auxiliary hyperplane ${\mathcal Z}_0$ at
$\tau = x=$const. (straight line). The curved hypersurface ${\mathcal Z}(\tau,x,y)$, which consists of free domains,
including a horizontal domain $AB$ near the origin,
approaches  the hyperplane ${\mathcal Z}_0$ at large distances. }}
\label{saconep}
\end{figure}

The presence of the flux $\partial \sigma$ is what causes deviation from sphericity.   The domain $ {\mathcal M}$ will be elongated (Fig.~\ref{saconep}).
A crucial question is whether this elongation
is in the $\vec n$  direction or is perpendicular to $\vec n$?.
 There are a few arguments
one can give supporting the first option: that
the exact solution of (\ref{balance}) will be elongated along the direction of $\vec n$, see 
Fig.~\ref{saconep}.
Consider the spherical solution as a starting point. 
Then the value of $\partial \sigma$ in  the direction
orthogonal to the sphere is $n \cos \theta$
where the angle $\theta$ is measured from the
vertical axis. This means that the north and south poles feel an outward force, while at the equator the extra force
vanishes.

Let us look at the same question from a slightly different perspective.
Let us call $r_1$ the ``radius" of ${\mathcal M}$ perpendicular to the $\vec n$
direction, and $r_2$ in the parallel direction.
Then the volume of ${\mathcal M}$ scales as
$V({\mathcal M})  \sim r_1^2 r_2\,.$
If $r_1 < r_2$ then ${\mathcal M}$  is elongated in the $y$ direction,
otherwise the elongation is in the perpendicular direction. In the search of solution we
may first find the shape for the $z$ field. The problem
is analogous to that from electrostatics,
of a conducting object ${\mathcal M}$ at a certain potential. The charges in general are  denser near the domain of strong curvature, implying that
$\partial z$ is larger at the poles in the case of   vertical 
orientation of ${\mathcal M}$ (Fig.~\ref{saconep})  or at the equator in the case of perpendicular orientation.
The larger the deviation from sphericity, the larger is the agglomeration of charges near the curved areas.
Now let us consider the sigma-field solution.  This case is different:
$\partial  \sigma$  always vanishes at the equator and is always maximal at the poles, independently of the shape.
The balance of  tensions on  the surface $\partial{\mathcal M}$ requires that, where $\partial  \sigma$ is larger, the slope $\partial z$ must also be larger to achieve compensation,
implying $r_1<r_2$.

Finally, we can try various particular ans\"atze for $\sigma (\tau, x, y)$
compatible with the asymptotic behavior (\ref{24}). For instance, if we
take\,\footnote{We assume $c$ and $a$ to be nonvanishing.
The ansatz (\ref{26}) contains $l=1$ and all higher odd waves
in a certain combination.}
\beq
\sigma_0 = n^\mu x^\mu + c\left(\frac{1}{|x + a|} - \frac{1}{|x - a|}
\right),
\label{26}
\eeq
where $c$ is a numerical coefficient and $a$ is a vector $a^\mu= \{a_0, a_1, a_2\}$,
then $\partial {\mathcal M}$ on which $\sigma_0$ vanishes exists
only if the vectors $n$ and $a$ are parallel, and then ${\mathcal M}$
is elongated along $n$ with necessity. 

The direction of elongation of ${\mathcal M}$ will be crucial
in what follows. Summarizing, we think it is fair to say
that the arguments presented above are compelling, although stop short
of proving the statement. To be  cautious,
for the time being, we will accept it as a motivated 
assumption.

What can be said of the bounce action under the above  conditions?
Let us first consider the contribution coming from $x^\mu \in {\mathcal M}$.
Given that  the composite brane is flat inside ${\mathcal M}, $ and $\sigma =0$
on the composite brane,
we get 
\beqn
\Delta S_{<}
 &=&
  -V({\mathcal M}) \left(2T_1 +\frac{e^2}{4\pi^2}\,n^2 -T_2\right),
  \nonumber\\[3mm]
  &\equiv&
-V({\mathcal M}) \Delta T + \left(\Delta S_\sigma\right)_{<}
\label{elevenpp}
\eeqn
where $V({\mathcal M})$ is the volume of ${\mathcal M}$, and the subscript $<$
indicates integration over $x^\mu \in {\mathcal M}$.

\begin{figure}[h!t]
\epsfxsize=6cm
\centerline{\epsfbox{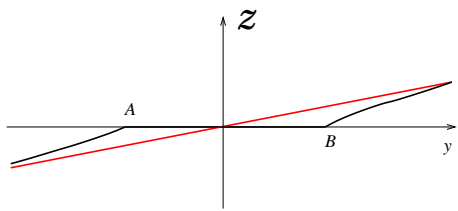}}
\caption{{\footnotesize A slice of the auxiliary hyperplane ${\mathcal Z}_0$ at
$\tau = x=$const. (straight line). The curved hypersurface ${\mathcal Z}(\tau,x,y)$, which consists of free domains,
including a horizontal domain $AB$ near the origin,
approaches  the hyperplane ${\mathcal Z}_0$ at large distances. }}
\label{sacone}
\end{figure}

Now, we have to calculate the loss $\Delta S_{>}$ coming from integration over
$x^\mu\,\, {\slash\!\!\!\!\!\in} \,{\mathcal M}$.
At first let us deal with $\left(\Delta S_\sigma\right)_{>}$,
\beq
\left(\Delta S_\sigma\right)_{>} =\frac{e^2}{4\pi^2}\,\int_>\, d^3 x \big[
\left(\partial\sigma\right)^2 -n^2\big]\,.
\eeq
Our task is to prove that $\left(\Delta S_\sigma\right)_{>}$ is larger than the absolute value
of the flux-related gain,
\beq
-\left(\Delta S_\sigma\right)_{<} =\frac{e^2}{4\pi^2}\,\int_<\, d^3 x  \,\,
n^2\,.
\eeq
This requirement is identical to the condition
\beq
\frac{e^2}{4\pi^2}\,\int \, d^3 x \big[
\left(\partial\sigma\right)^2 -n^2\big] >0\,,
\label{28}
\eeq
where the integral runs over the entire three-dimensional space.
The easiest way to see that Eq.~(\ref{28}) is satisfied
is through an auxiliary geometrical picture. Indeed, let us introduce an auxiliary coordinate ${{\mathcal Z}\perp \tau,x,y}$
such that 
\beq
{\mathcal Z} = \frac{\sigma}{\mu }\,,\qquad {\mathcal Z}_0 = \frac{n\,y}{\mu}\,,
\eeq
where $\mu$ is an auxiliary (large) parameter of dimension of mass. Then ${\mathcal Z}_0$
represents a slightly tilted three-dimensional hyperplane, see Fig.~\ref{sacone}.
The integral in Eq.~(\ref{28}) represents $\smallfrac{e^2\mu^2}{4\pi^2}\times$the
difference between the hyperareas (volumes)  of the curved and flat three-dimensional  hypersurfaces
(their $\tau=x=$const. slice is shown in Fig.~\ref{sacone}). Needless to say, considering  $AB$ alone 
we would get a negative
contribution. However, overall, the curved  hypersurface has a larger area than the flat one
given that both hypersurfaces touch each other at $r=\infty$.
This completes the proof of the most crucial statement -- the total flux-related contribution in $S_B$ is always positive, as was the case in the small-flux limit
(Section~\ref{small}).

Various estimates we have carried out for reasonably shaped probe
${\mathcal M}$'s show that
\beq
\Delta S_\sigma =\kappa_{{\mathcal M}}\, V({\mathcal M}) \, \frac{e^2\,n^2}{4\pi^2}
\label{30}
\eeq
where $\kappa_{{\mathcal M}}$ is a positive number depending in geometry of ${\mathcal M}$.
Moreover, if ${\mathcal M}$ is elongated in the $n$ direction, as was argued above, 
$\kappa_{{\mathcal M}} > 1$. Although we were unable to obtain a general proof of this inequality,
it is quite transparent. The spherical shape has $\kappa_{{\cal S}^2} =1$. The previous argument showed that the actual solution is elongated along the $n^{\mu}$ direction.
Given the same volume of ${\mathcal M}$, elongation in the direction of $n$
will create a larger disturbance (see Fig.~\ref{sacone}), 
 increasing the coefficient $\kappa$. 
 
 It might be instructive to consider a particular example.
 Consider $\sigma$ comprised of $l=1$  and $l=3$ harmonics,
\beq
\sigma = \cos\theta \left( r - \frac{1}{r^2}\right) + \alpha \frac{5 \cos^3\theta - 3 \cos\theta}{ r^4}\,,\qquad \cos\theta = \frac{n^\mu x^\mu}{|n|\cdot|x|}\,,
\eeq
where
$\alpha$ is a coefficient that deviates the surface $\sigma=0$ from sphericity.
 Positive or negative $\alpha$ implies elongation of ${\mathcal M}$ in the $vec n$ direction or perpendicular.
The flux-related contribution to the bounce action is independent of $\alpha$.
On the other hand,  the volume of of ${\mathcal M}$ does depend on it.
The volume  $V({\mathcal M})$ becomes smaller in the  case 
of parallel orientation (Fig.~\ref{saconep}) and larger in the perpendicular case.
 
The fact that $\kappa_{{\mathcal M}}$ is  positive can be seen from an alternative argument.
Indeed, let us  use the  multipole expansion for the $\sigma$ field at large $r$.
The expansion then takes the form
\beq
\sigma = n^\mu x^\mu \left(1 - \frac{p_{\cal M}}{r^3}\right) +  O \left( \frac{1}{r^4}\right) ,
\label{eight}
\eeq
where the coefficient $p_{\mathcal M}$ presents the dipole term, 
while all higher multipoles are hidden in $O(1/r^4)$ terms. It is natural
to expect $p_{\mathcal M}$
to be positive.
The flux-related contribution to the bounce action  is 
\beqn
\Delta S_{\sigma} &=& \frac{{e}^2}{4\pi^2}    \int   d^3x\, 
\Big\{\left(\partial^\mu\sigma\right)^2 -
\left[\partial^\mu\left(n^\alpha x^\alpha\right)
\right]^2\Big\}
\nonumber\\[3mm]
&=&
\frac{{e}^2}{4\pi^2}\,
\int_{S_2} d S^\mu\Big[\sigma\partial^\mu\sigma -\left(n^\alpha x^\alpha\right) n^\mu
\Big]
\label{34}
\eeqn
where we performed integration by parts and used the fact that $\Delta \sigma =0$.
The second integral in Eq. (\ref{34}) runs over the surface of the large 
sphere,\footnote{Strictly speaking, $\partial \sigma$ is not continuous on $\partial
{\mathcal M}$. The discontinuity should be smoothed out. This produces no impact
in Eq. (\ref{34}).} $S_2(R\to\infty )$. From Eq. (\ref{34}) it is
clearly seen that only the dipole term in $\sigma$ contributes to $\Delta S_{\sigma}$,
all other multipoles fall off at infinity too fast to contribute. The result is
\beq
\Delta S_{\sigma} =p_{\mathcal M} \frac{e^2\,n^2}{3\pi}\,.
\eeq
Comparing with Eq.~(\ref{30}) we conclude that
\beq
 p_{\cal M} = \frac{3}{4\pi}\, \kappa_{\mathcal M}\,  V({\mathcal M})  \,.
\eeq
To compute exactly the critical value of flux we need the relation between $p_{\cal M}$ and $V$ (i.e. the coefficient $\kappa_{\mathcal M}$). As was discussed above, 
 $\kappa_{\mathcal M} \geq1$ is to be expected.

Our next task is to analyze $\Delta S_z$. The value of $\left(\Delta S_z\right)_<$
can be read off from Eq.~(\ref{elevenpp}).
\beq
\left(\Delta S_z\right)_< = -V({\mathcal M}) \,\,\Delta T \,.
\eeq
This is the only negative contribution (gain) in $S_B$.
The value of $\left(\Delta S_z\right)_>$
can be determined as follows.
At $x^\mu\,\, {\slash\!\!\!\!\!\in} \,{\mathcal M}$ the solution for $z(\tau,x,y)$ can
 be expanded in spherical 
harmonics, starting from $l=0$. Given the boundary conditions for $z$ 
this multipole expansion at large $r$ takes the form
\beq
z= \frac{d}{2} - \frac{q}{r} + {\rm higher \,\, harmonics} \,.
\label{eightp}
\eeq
All angular dependence resides in higher harmonics.
The $l=2k$ harmonics are suppressed by $1/r^{1+2k}$ where $k=1,2,...$.
The coefficient $q$ 
depends on details of the bounce solution, and in particular, on geometry of
${\mathcal M}$, i.e. $q=q_{\mathcal M}$. What is important for us in what follows  is that
\beq
q_{\mathcal M} >0\,.
\eeq
Moreover, $q_{\mathcal M} $ has dimension  $[m]^{-2}$  and is of the order of $d\,\ell_{\mathcal M} $
(cf. the small flux limit in which $q_{\mathcal M} =r_*\,d/2$).
Then $\left(\Delta S_z\right)_>$
 can be expressed uniquely as a function of this coefficient. Indeed, integrating by parts and using the
 fact that $z$  is a harmonic function at $x^\mu\,\, {\slash\!\!\!\!\!\in} \,{\mathcal M}$ which vanishes at
 $\partial{\mathcal M}$ we get
 \beqn
 \left(\Delta S_z\right)_> &=& T_1\,\int_> \,d^3 x\,  (\partial z)^2 = T_1\left[\int_{\partial {\mathcal M}}\, 
 \partial^\mu\left(z\partial^\mu z\right) +
 \int_{\infty}\, 
 \partial^\mu\left(z\partial^\mu z\right)\right]
 \nonumber\\[3mm]
&- &
  T_1\,\int_> \,d^3 x\,  z\,\Delta z = 2\pi\,T_1\,q_{\mathcal M}\, d\,.
 \eeqn
 The only nonvanishing contribution to $ \left(\Delta S_z\right)_>$ comes from the
 surface
 integral $\int_{\infty}\, 
 \partial^\mu\left(z\partial^\mu z\right)$ and only from the $l=0$ harmonics
 in Eq.~(\ref{eightp}). All higher harmonics fall off too fast at infinity and
 do not affect the above surface integral. As a result,
 \beq
 \Delta S_z =2\pi\,T_1\,q_{\mathcal M}\, d -V({\mathcal M}) \,\,\Delta T \,.
 \eeq

The bounce action is determined by one extra positive parameter $\kappa_{\mathcal M}$
(see Eq. (\ref{30})) which at the moment is not yet firmly established,
\beq
  \Delta S_\sigma +  \Delta S_z =\kappa_{{\mathcal M}}\, V({\mathcal M}) \, \frac{e^2\,n^2}{4\pi^2}
-V({\mathcal M}) \,\,\Delta T +2\pi\,T_1\,q_{\mathcal M} \, .
\label{36}
\eeq
The quantum fusion occurs only if $ \Delta T >  {\kappa_{{\mathcal M}}/e^2n^2}{4\pi^2} $.
 This determines a critical value of $\Delta T$ below which the domain wall fusion is
 impossible,
 \beq
\Delta T_* = \frac{\kappa_{{\mathcal M}}e^2n^2}{4\pi^2}\,.
\eeq
 To find the critical bounce action we have to extremize
(\ref{36}) with respect to the bounce size. What we can do is to find the extremum with regards
to dilatations of $x_\mu$. To this end we take into account that 
$ V({\mathcal M})$ scales as $\ell_{\mathcal M}^{\,\,3}$ 
while $q_{\mathcal M} \sim \ell_{\mathcal M} d$. From this we conclude that
\beq
S_B \geq \frac{4\pi\,T_1\,q_{\mathcal M}}{3}\,.
\eeq
At $\Delta T < \Delta T_*$ no balance between gain and loss in $S_B$
is achievable. At $\Delta T = \Delta T_*$ the coefficient $q_{\mathcal M}$
must tend to infinity.

\section{Conclusions}

In the small flux limit we explicitly find  the bounce solution and  the
fusion rate as a function of the flux. We argue that at large (antiparallel) fluxes there exists
 a critical value of the flux (versus $2T_1-T_2$), which switches off quantum fusion altogether.
 However, a reservation is in order here:
our consideration of the  flux-related wall stabilization
is based on substantiated arguments that, nevertheless,  stop short of unquestionable proof.
 We used the same framework as in Ref.~\cite{Bolognesi:2009kp}.
 In particular, the binding energy is assumed to be much less than the wall tension,
 so that the DBI action can be expanded up to quadratic in derivative terms,
 while higher terms neglected.
 
 \vspace{7mm}
 We are grateful to Adi Armoni and Sasha Gorsky for discussions and stimulating questions.
 This work  is supported in part by DOE grant DE-FG02-94ER408. 


\end{document}